\documentclass[12pt,aps,prd,showpacs,floatfix,nofootinbib]{revtex4-2}
\usepackage[utf8]{inputenc}

\usepackage{color}
\usepackage{graphicx}
\usepackage{bm}
\usepackage{amsmath}
\usepackage{amsfonts}
\usepackage{eufrak}
\usepackage{hyperref}
\usepackage{natbib}
\usepackage{lineno}

\newcommand{\mathbit}[1]{\boldsymbol{#1}}

\begin{document}
	\title{Two Dimensional Turbulence in a Massless Fluid with a Relativistic Lattice Boltzmann Model}

	\author{Mark Watson}

	\affiliation{Department of Physics, University of Colorado, Colorado Springs, Colorado 80904, USA  }
	
	\begin{abstract}	
		We investigate a relativistic adaptation of the Lattice Boltzmann Method that reproduces the equations of motion for a turbulent, two-dimensional, massless hydrodynamic system. The classical Lattice Boltzmann Method and its extension to relativistic fluid dynamics is described.  The numeric formulation is evaluated using a zero-averaged stirring force introduced into the numerics to induce turbulence, and the flow characteristics produced are compared to properties of a classical turbulent hydrodynamic flow.  The model can reasonably be expected to offer quantitative simulations of electron fluid flows in graphene or Kagome lattices.
	\end{abstract}
	
	\maketitle
	
	\section{Introduction}
	
The study of the phenomenon of turbulence in hydrodynamic systems is vital in understanding the dynamics of fundamental systems in the physical world, from atmospheric weather patterns to astronomical systems.  Fluid systems with relativistic flows such as the jets that accompany compact astrophysical objects like neutron stars and black holes, or those modeling heavy ion collisions \cite{rezzolla_zanotti_2013} are of interest for their unique hydrodynamic characteristics.  The motion of a system of massless particles can be viewed as a relativistic fluid as the dynamics of the individual fermions are described by the Dirac equation.  When the particles are considered continuously they form a fermionic fluid governed by relativistic hydrodynamics.  It is thought that the flow of the charge carriers in a two-dimensional solid lattice formed by graphene may be considered a massless, relativistic, viscous electronic fluid \cite{2015PhRvB..91h5401F} \cite{Gabbana_2018}.  The $sp^2$ hybridized orbit of the electrons in the atoms of the hexagonal lattice form resilient $\sigma$ bonds with neighboring atoms that create a linear dispersion relation forming so-called Dirac cones in the energy spectrum.  Low energy excitations of the particles and the particle holes create massless charge carriers in the lattice whose motion follows Dirac's equation; Dirac fermions \cite{2005Natur.438..197N}.  Near the points of the cones these charge carriers form a massless electronic fluid where the Fermi velocity $v_F$ $(\sim 10^6 \frac{m}{s})$ plays the role of the speed of light.  The fluid velocities, bounded from above by $v_F$, are found to be $0.3$ percent the speed of light.  Despite these non-relativistic velocities, the electron fluid behaves very much like a relativistic fluid since there are no mass-scales in the problem.  For this reason, it is expected that the electron velocity probability distribution follows the relativistic J{\"u}ttner distribution instead of the classical Maxwell distribution.  This niche (non-relativistic velocities yet relativistic velocity distribution) is the focus of the present work.  Due to the smaller velocity, the flow in the electronic fluid in graphene is likely pre-turbulent or non-turbulent.  Recently, there has been interest in the lattice structure of a solid known as a Kagome solid that also exhibits a linear dispersion relation in its energy spectrum that forms Dirac cones \cite{2019arXiv191106810D}.  This lattice structure appears to support a stronger coupling constant estimated to be $3.2$ times that of graphene due to the orbital hybridization.  The bulk viscosity caused by phonons and impurities negligible, the dominant coulomb forces drive a larger fluid velocity and a simultaneously smaller shear viscosity.  These properties allow the emergence of a turbulent massless fluid that is accessible via experiment.  The work by Di Sante et al. \cite{2019arXiv191106810D} uses a holographic approach to determine the shear viscosity of the electronic fluid in a Kagome solid and was able to measure it experimentally.  


Due to the non-linearity of the equations of motion, particularly in the turbulent regime, the study of hydrodynamics often requires a numeric approach to find solutions and many different computational methods have been employed.  The wide variety of methodologies include mathematical modeling of the Euler, Navier-Stokes, and Boltzmann equations, and modeling with particle-based solvers.  Each has its advantages and disadvantages.  

The Lattice Boltzmann Method (LBM), born out of a particle-based solver and based on the Boltzmann equation, is one such approach and has had a great deal of success in the modeling of hydrodynamics over the last few decades \cite{PhysRev.94.511}\cite{doi:10.1146/annurev.fluid.30.1.329}\cite{succi2001lattice}.  It is conceptually simple, highly parallelizable, and unconditionally stable in the presence of shocks owing to its reliance on Boltzmann's probability distribution function.  

LBM modelers of relativistic hydrodynamic flow are less abundant. A Relativistic Lattice Boltzmann Method (RLBM) described by Mendoza, Boghosian, Hermann and Succi \cite{2010PhRvD..82j5008M} uses parameter matching against conservation laws to fix the form of the equilibrium probability distribution function used in the collision operation. Similar schemes for ultra relativistic flows have been proposed by Mohseni, Mendoza, Succi and Herrmann \cite{2013PhRvD..87h3003M}, and Mendoza, Karlin, Succi and Herrmann \cite{2013PhRvD..87f5027M}.  A full description of a dimension-independent procedure to design a relativistic lattice Boltzmann scheme is presented by Gabbana, Simeoni, Succi and Tripiccione \cite{2020PhR...863....1G}.  A variation of the RLBM proposed by Romatschke, Mendoza and Succi (2011) \cite{2011PhRvC..84c4903R} expands the J{\"u}ttner distribution as the equilibrium probability distribution function.  An extension of this work, applicable when the chemical potential is not constant, is described by Ambrus and Blaga \cite{2018PhRvC..98c5201A}.  The present work extends \cite{2011PhRvC..84c4903R} to model turbulent flows by including a forcing term in the relativistic lattice Boltzmann algorithm.  With sufficient viscosity in the collision operator and with large enough forcing magnitudes, this term produces Reynolds numbers in the simulation sufficient to create flows in the turbulent regime.  The model is validated through numerical comparison against established turbulent hydrodynamic theory.  Using this proposed relativistic turbulent hydrodynamic solver we wish to explore the characteristics of turbulent massless fluid flow.

    \section{Lattice Boltzmann Method}  \label{section:stdLBM}

Some explanation of the standard/classical Lattice Boltzmann Method may be appropriate.  The standard LBM grew out of the Lattice Gas model, which is a particle-based solver that models particles moving and colliding within a discrete lattice framework.  The LBM approach replaces the particles in the lattice with the conceptualized 6 + 1 dimensional probability distribution function $f(t, \mathbit{x}, \mathbit{v})$, quantifying the probability that a particle will be in a given state with regard to position and velocity at a given time.  The dynamics of the probability distribution function are described by the well-known Boltzmann equation.  
\begin{equation} \label{eq:boltzmaneq_cl}
\left[ \partial_t + \mathbit{v} \cdot \nabla - \frac{1}{m} \mathbit{F} \cdot \nabla^{ (v) }  \right] f = C \left[ f \right]
\end{equation}
$\mathbit{v}$ and $m$ represent the particle's velocity and mass, $\mathbit{F}$ is an external force, and $C \left[ f \right]$ is the collision operator, originally described by Boltzmann as a complex six-dimensional integral.  Boltzmann's form of the operator is difficult to model computationally so it is commonly replaced by a simpler Bhatnagar-Gross-Krook (BGK) type collision operator \cite{1954PhRv...94..511B}, 
\begin{equation} \label{eq:bgk_cl}
C \left[ f \right] = \frac{f-f_{eq}}{\tau_R}\text{.}
\end{equation}
In this operator $f_{eq}$ is the local probability distribution function at equilibrium, and $\tau_R$ is the time to relaxation to the local state of equilibrium.  This simplified collision operator ansatz sacrifices the complex microscopic details of the particle interactions, but faithfully preserves the evolution of the fluid's macroscopic moments by maintaining conservation of mass, momentum and energy.

The macroscopic moments, specifically the local mass density $\rho$, the fluid velocity $ \mathbit{u} $, and the total (kinetic plus internal) energy $\epsilon $ are found through integration of the probability distribution function over velocity space \cite{2016PhRvA..93a3618B}.
\begin{align} \label{eq:moments_cl}
\nonumber \rho             \equiv & m           \int d^3\mathbit{v}\ f, \\ 
\nonumber \rho \mathbit{u} \equiv & m           \int d^3\mathbit{v}\ \mathbit{v} f, \\
          \epsilon         \equiv & \frac{m}{2} \int d^3\mathbit{v}\ |v|^{2} f
\end{align}
These definitions also hold true when the flow is at a local equilibrium state.  Then the probability distribution function is the equilibrium probability distribution function; $f \rightarrow f^{eq}$.  The equilibrium probability distribution function can be determined in terms of the macroscopic moments.  Consider a fluid at equilibrium, when its microscopic velocity is isotropic in velocity space and is therefore equivalent to the macroscopic velocity $\mathbit{u}$ ($\mathbit{v}=\mathbit{u}$), and can be considered to be constant ($|v|^2= \rm{const}$).  Following the same lines of Maxwell's derivation of the equilibrium distribution function \cite{doi:10.1080/14786446008642818} and using the definitions in (\ref{eq:moments_cl}), it is straightforward to find the form of the equilibrium probability distribution function in terms of the macroscopic moments:
\begin{equation} \label{eq:eq_dist_funct_cl}
f^{eq}(t, \mathbit{x}, \mathbit{v}) = 
   \frac{\rho}{m} 
   \left( \frac{1}{ 2 c_s^2 (T) \pi } \right) ^{\frac{3}{2}} 
   e^{ \frac{ -\left| \mathbit{v} - \mathbit{u} \right|^2}{ 2 c_s^2 (T) } }
\end{equation}
This form introduces the square of the local sound speed as a function of temperature $T$, $c_s^2 (T) = \frac{T}{m}$.  This definition of the equilibrium distribution function is used within the BGK collision operator to complete the equations for the evolution of the probability distribution function.

The sacrifice of collision details in favor of the macroscopic moments permits the dynamics of the system to be modeled accurately and efficiently based on the moments with an appropriately defined computational model.  The Lattice Boltzmann Method is a family of modeling methods employed to model the dynamics of the probability distribution function through the Boltzmann equation projected onto a discrete spatial lattice with a superimposed discrete velocity space.  This provides the model with the minimum amount of information required to solve the Boltzmann equation accurately.  There is an abundance of literature on the implementation and application of LBM models; see for instance \cite{succi2001lattice} and \cite{kruger_lattice_2017}.  

In an LBM model, space is discretized into a cubic lattice where all the information in the probability distribution function for a discrete volume is contained within the nodes of the lattice, separated from each other by a lattice spacing in one, two, or three dimensions.  The velocity space is also discrete at every spatial lattice node using a quadrature analysis method that identifies only the discrete speeds necessary to preserve the moments.  In that way, the LBM lattice can be considered a velocity lattice.  

 
A closer look at how a discrete velocity space is determined using a quadrature technique provides insight into the LBM mechanism.  Given a function $f(x)$, expressed as $f(x) = \omega(x)g(x)$, where $\omega(x)$ is an appropriately chosen weighting function, an integral of the form $ \int_{-\infty}^{\infty} \omega (x) g(x) dx $ can be approximated by a summation of the function $g(x)$ evaluated at a suitable choice of abscissae $x_i$ and a suitable choice of weights $w_i$ where $i=1...n$ \cite{abramowitzstegun}.
\begin{equation} \nonumber
\int_{-\infty}^{\infty} \omega (x) g(x) dx \approx \sum_{i=1}^{n} w_i g(x_i)
\end{equation}
If $g(x)$ is represented using a class of orthogonal polynomials, the weighting function $\omega(x)$ is the generating function associated with the polynomial set, and the abscissa $x_i$ are the $n$ roots of the polynomials.  Furthermore, if the function $g(x)$ is of order $N$ or less, then using $n$ orthogonal polynomials evaluates the integral \textit{exactly} when $N \le 2n-1$.
\begin{equation} \label{eq:gauss_quad_with_ortho_polys}
\int_{-\infty}^{\infty} \omega (x) P^{(N)}(x) dx = \sum_{i=1}^{n} w_i P^{(N)}(x_i)
\end{equation}
Here the set of functions $P^{(N)}$ are the chosen orthogonal polynomials up to order $N$, and the weights $w_i$ are determined based on the choice of polynomials.

The limits of integration in the integral determine which set of orthogonal polynomials are best to use.  For an integration over $[-1,1]$, Legendre polynomials are found to be useful.  For the range $\left[0,\infty\right)$, Laguerre polynomials are best suited.  And, for the integral limits described in (\ref{eq:gauss_quad_with_ortho_polys}), $(-\infty,\infty)$, Hermite polynomials are the appropriate choice.  We note that the equilibrium probability distribution function in (\ref{eq:eq_dist_funct_cl}) used in the Boltzmann equation is of the same form as the generating function of the Hermite polynomials.
\begin{equation} \nonumber
H_n (x) = (-1)^n \omega (x) \frac{d^n}{dx^n} ( \omega (x) )^{-1}, \ \ \ \omega (x) = e^{x^2}
\end{equation}
This implies the straightforward use of an expansion of the equilibrium probability distribution function in terms of Hermite polynomials, and the use of the Gauss-Hermite quadrature integral estimation technique.
\begin{equation}
\int_{-\infty}^{\infty} \omega (x) H^{\left(N\right)}\left(x \right) dx \approx \sum_{i=1}^{n} w_i H^{\left(N\right)}\left(x_i \right)
\end{equation}
In this case the discrete weights are determined by:	
\begin{equation} \nonumber
w_i=\frac{n!}{\left(nH^{\left(n-1\right)}\left(x_i\right)\right)^2}
\end{equation}
The Gauss-Hermite quadrature integral relation is generalized to $d$ dimensions as \cite{kruger_lattice_2017}
\begin{equation}
\int_{-\infty}^{\infty}{d^dx\ \omega\left(\mathbit{x}\right)\mathbit{H}^{N}\left(\mathbit{x}\right)}=\sum_{i=1}^{n^d}{w_i\mathbit{H}^{N}\left(\mathbit{x}_i\right)}
\end{equation}
This quadrature technique can be used in a straightforward way to evaluate the velocity space integrals (\ref{eq:moments_cl}) using a Hermite polynomial expansion of the probability distribution function.
\begin{align}
\nonumber \rho             =& m           \sum_{i} f_i, \\
\nonumber \rho u_{\alpha}  =& m           \sum_{i} f_i v_{i\alpha},\\
          \epsilon         =& \frac{m}{2} \sum_{i} f_i v_{i\alpha}v_{i\beta} = \rho c_s^2 \delta_{\alpha \beta} + \rho u_{\alpha} u_{\beta}
\end{align}
In these summations $f_i$ expands $f$ about a velocity evaluated at the abscissae, and incorporates the weights.  A discrete summation to evaluate a continuous integral implies there is a finite number of velocities $n$ that are needed to faithfully reproduce the fluid macroscopic moments (\ref{eq:moments_cl}).  That is, the roots of the three-dimensional Hermite polynomials, the abscissae $x_i$, determine the discrete velocities that are required to connect the nodes of the lattice.  The number of abscissae chosen need only be enough to match the order of the highest moment, $N=2n-1 \Rightarrow n=\frac{N-1}{2}$.  Different LBM formulations use different numbers of speeds depending on precision and computational efficiency requirements.  $n = 19$ and $27$ are commonly used for a 3 dimensional lattice in configurations named D3Q19 and D3Q27, respectively.  

\begin{figure}[h]
	\includegraphics[width=0.5\linewidth]{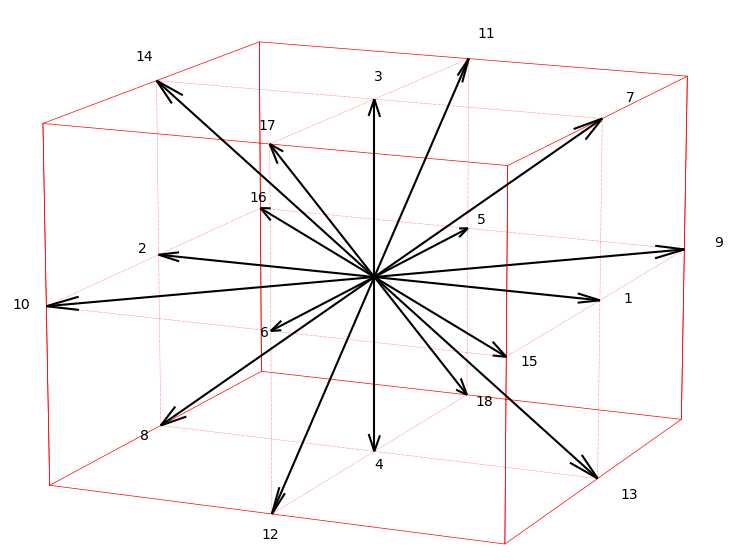}
	\caption{Single node of a D3Q19 configured lattice.  The node at the center of the cube is connected to its neighboring nodes by 19 discrete speeds (including 0).}
	\label{fig:d3q19cube}
\end{figure}

Using this discretization formula, the continuous probability distribution function is then re-expressed in discrete form: $ f(t,x,v) \rightarrow f_i(t,x_j) $, where $f_i$ is the probability distribution function at lattice node $x_j$ for the discrete speed $i$, incorporating the weighting factor $w_i$. The Lattice Boltzmann Equation is then:
\begin{equation} \label{eq:lbe_cl}
f_i \left(t + \Delta t, x_j + v_i \Delta t \right) - f_i \left( t, x_j \right) = \Delta t C \left[ f_i \left( t, x_j \right) \right]
\end{equation}
Figure \ref{fig:d3q19cube} depicts a single node of a three dimensional lattice with 19 speeds.  Each node contains 19 different probability distribution function values $f_i$, one for each speed $v_i$ at that lattice location.

The principal of dynamic similarity asserts that the fluid equations of motion are the same at all length scales, so it is customary within a LBM formulation to rescale the space and time values to dimensionless quantities with respect to some length $R_{\perp}$ and frequency $\omega_{\perp}$ [\cite{2016PhRvA..93a3618B}].
\begin{equation}
\bar{\mathbit{x}} = \frac{\mathbit{x}}{R_{\perp}} ,\ \ \bar{t} = t \omega_{\perp},\ \ 
\bar{\mathbit{v}} = \frac{\mathbit{v}}{R_{\perp} \omega_{\perp}} ,\ \ \bar{\mathbit{u}} = \frac{\mathbit{u}}{R_{\perp} \omega_{\perp}} 
\end{equation}
The rescaled quantities are then easily converted to the scale of the domain being modeled when they are recorded.
With both Euclidean space and velocity space discretized, the Lattice Boltzmann Method then simulates the time evolution of a hydrodynamic system with discrete time steps following this process: 
\begin{enumerate}
	\item Initialize the local equilibrium distribution function $f_i^{eq}$ at each node on the grid according to defined initial conditions of the macroscopic variables, and set the new local probability distribution function to be equal to the equilibrium probability distribution function, $f_i^*=f_i^{eq}$. 
	\item \label{list:lbmproc_init_cl} “Stream” the new distribution function $f_i^*$ to the neighboring node connected by the discrete speed, accounting for boundary conditions.
	\item Calculate the macroscopic moments based on the local state at each node, optionally recording them for analysis.  
	\item Calculate the equilibrium probability distribution function $f_i^{eq}$ based on the moments.
	\item Perform the collision calculation locally using the equilibrium probability distribution function $f_i^{eq}$ and the local probability distribution function $f_i$ to find the new probability distribution function:  $f_i^* = f_i + \Delta t C $.
	\item Repeat from step \ref{list:lbmproc_init_cl} for each time step in the simulation.
\end{enumerate}

This technique is advantageous computationally since, though the dynamics of a hydrodynamic system are non-linear (see (\ref{eq:boltzmaneq_cl})), the non-linear computations are carried out locally at each node independently and then streamed linearly to the (non-local) neighboring nodes. This allows each node to be computed independently and simultaneously with all the other nodes, providing a high degree of parallelism.  One has only to ensure the simpler streaming step happens serially after the collision calculation for all nodes has been completed.
    
    \section{Relativistic Lattice Boltzmann}
    
    \subsection{Relativistic Boltzmann}
    
The classical version of the LBM has seen wide adoption for the simulation of a variety of hydrodynamic systems.  Many variants have been adopted to improve accuracy or performance or to address other difficulties that can arise in different domains.  One such variation proposed by Romatschke, Mendoza and Succi in 2011 \cite{2011PhRvC..84c4903R} adjusts the Lattice Boltzmann Method framework to model relativistic hydrodynamic flows.  This Relativistic Lattice Boltzmann Method (RLBM) adaptation is based on the dynamics of the relativistic version of the probability distribution function, $ f=f( x^\mu, p^\nu) $, where $x^\mu$ is the position 4-vector and $p^\nu$ is the momentum 4-vector.  The dynamics of the probability distribution function are described by the relativistic Boltzmann equation with a relativistic analog of the BGK collision term \cite{2002rbet.book.....C}.
\begin{equation} \label{eq:boltzmanneq_rel}
\left[ p^\mu \nabla_\mu - \Gamma_{\mu\nu}^\lambda p^\mu p^\nu  \partial_{\lambda}^{(p)} \right] f = 
- \frac{p^\mu u_\mu }{ \tau_R } \left( f - f_J^{eq}  \right)
\end{equation}
For this equation, and for the remainder of this treatment, the units are natural; $ c=k_B=\hbar=1$.  $ \nabla_\mu $ is the covariant derivative, and $\Gamma_{\mu\nu}^\lambda$ is the Christoffel symbols which are given by derivatives of the underlying metric tensor $g_{\mu \nu}$.  For this treatment we assume the Minkowski metric for a flat space-time configuration.  The macroscopic fluid velocity is $ u^\nu = \gamma \left( 1, v^i  \right) $, where $ \gamma = \left( 1 - v^2 \right)^{-\frac{1}{2}} $ is the Lorentz factor and $ v^i $ is the microscopic particle 3-velocity.  In this description, Greek indices refer to 4-vector space-time components, and Latin indices are used for 3-vector spatial components.  In the relativistic regime energy, $ E = \sqrt{m^2 + p^2}$, is no longer quadratic in the velocity.  This is reflected in the relativistic form of the equilibrium probability distribution function $f_J^{eq}$ which is the J{\"u}ttner distribution.
\begin{equation} \label{eq:equlibriumdf_rel}
f_J^{eq}=Z^{-1} e^{ \frac{-p_\mu u^\mu }{T} }
\end{equation}
Here $T$ is the local temperature and $Z$ parameterizes the number of degrees of freedom, which in this work will be taken to be one. 

The macroscopic moments in relativistic hydrodynamics are obtained from the energy-momentum (energy-stress) tensor $T^{\mu \nu} $ which can be obtained by integration over the 4 components of momentum space \cite{2011PhRvC..84c4903R}.
\begin{align} \label{eq:energyMomentumIntegral}
\nonumber T^{\alpha \beta}  \equiv & \int d \chi p^\alpha p^\beta f(t, x, p) \\
\equiv & \int \frac{d^4p}{\left( 2 \pi \right)^{3}} \delta \left( p^\mu p_\mu - m^2 \right) 2 H \left( p^0 \right) p^\alpha p^\beta f(t, x, p)
\end{align} 
In this definition $m$ is the particle's mass and $H$ is the Heaviside step function.  (Note the notation difference for the relativistic probability distribution function, $f(x^\mu, p^\nu) \rightarrow f(t, x, p) $.)  Once obtained, the symmetric energy-momentum tensor contains the energy density ($T^{00}$), the momentum density ($T^{0i}$), and the momentum flux density ($T^{ik}$).  

Using the J{\"u}ttner form of the equilibrium probability distribution function, integration of (\ref{eq:boltzmanneq_rel}) at equilibrium produces the equilibrium energy-momentum tensor $T_{eq}^{\mu \nu}$ with the familiar relation
\begin{equation} \label{eq:stressEnergyT}
T_{eq}^{\mu \nu} = \left( \epsilon + P \right) u^\mu u^\nu + P g^{\mu \nu} 
\end{equation} 
where both energy density and pressure, $ \epsilon $ and $P$, are functions of temperature.  Then the viscosity coefficient, represented in the collision term, is  $ \eta = \tau_R \frac{\epsilon + P}{5} $, with relaxation time $ \tau_R = 5 \frac{\eta}{s} T^{-1} $, where $s$ is the entropy density \cite{2017arXiv171205815R}.  We apply the canonical, so-called Landau-Lifshitz condition, where $ u_\mu T^{\mu \nu} \equiv \epsilon u^\nu $, to provide a relationship between the energy density and the fluid velocity in the rest frame of reference.  Because the particles modeled are considered to be massless, the equation of state used is simply that of an ideal gas, $ P = \epsilon / 3 $.

	\subsection{Boltzmann to Lattice Boltzmann}
This continuous formulation of a relativistic hydrodynamic system is projected onto a discrete lattice for computational modeling following a prescription similar to the classical LBM.  Space and time $ x^\mu $ are discretized in the same manner, but the momentum $p^\mu$ must be addressed differently.  The relativistic form of the equilibrium distribution function is not quadratic in the exponent, and therefore has a different form than the Hermite polynomial generator function; $e^{-\frac{v^2}{T}}\ \rightarrow e^{-\frac{p^\mu u_\mu}{T}}$.  Hence, the discretization of the velocity space is less straight forward.  To address this, the equilibrium probability distribution function is re-expressed as an expansion about powers of $ \frac{ |\mathbit{p}| u^0 }{ T_0 \theta } $.
\begin{equation}
e^{- \frac{p \cdot u}{ T } } = 
e^{- \frac{ |p| u^0}{ T_0 \theta } } \sum_n 
\left( \frac{ \mathbit{p} }{ |\mathbit{p}| } \right)^n 
\left( \frac{ |\mathbit{p}|u^0 }{ T_0 \theta  } \right)^n
\frac{ \left( \mathbit{u} \right)^n  }{ \left( u^0  \right)^n n! }
\end{equation}
Here the momentum is expressed as unit vectors, $ \mathbit{v} = \frac{\mathbit{p}}{ |\mathbit{p}| }$, and $ \theta$ is the scaled temperature with respect to a reference temperature $T_0$, $ \theta = \frac{T}{T_0} $.  In this form we can express the relativistic probability distribution function in terms of a (different) set of orthogonal polynomials as:
\begin{equation} \label{eq:rel_pdf_exp}
f(t,x, p) = e^{ \frac{-p^0}{T_0} } \sum_n P_{ i_1 ... i_n }^{ (n) } \left( \mathbit{v} \right) a_{ i_1 ... i_n }^{ (n) } \left( t, x, \frac{p^0}{T_0} \right) 
\end{equation}
where $ p^0 = |\mathbit{p}| $.  The chosen polynomials $ P_{ i_1 ... i_n }^{ (n) } \left( \mathbit{v} \right) $ are three dimensional polynomials about the unit vector velocity determined by orthogonality with respect to the angular integral $ \int \frac{d\Omega}{4\pi} $.  Their properties are listed in appendix \ref{appendixA}.  The coefficients in (\ref{eq:rel_pdf_exp}) are determined using orthogonality conditions:
\begin{align} \label{eq:rel_pdf_exp_coef}
\nonumber a^{(0)} \left(t, x \right) 
	= \frac{1}{\left( \alpha \right)!} \int d\bar{p}\ \bar{p}^\alpha \int \frac{d\Omega}{4\pi} f P^{(0)} \left( \bar{p} \right) , \\
\nonumber a_i^{(1)} \left(t, x \right) 
	= \frac{3}{\left( \alpha \right)!} \int d\bar{p}\ \bar{p}^\alpha \int \frac{d\Omega}{4\pi} f P_i^{(1)} \left( \bar{p} \right) , \\
          a_{ij}^{(2)} \left(t, x \right) 
	= \frac{15}{2 \left( \alpha \right)!} \int d\bar{p}\ \bar{p}^\alpha \int \frac{d\Omega}{4\pi} f 	P_{ij}^{(2)} \left( \bar{p} \right)
\end{align}
where $\bar{p}=p^0/T_0$ is the scaled momentum magnitude.  The parameter $\alpha$ is chosen to be 3, reflective of the number of dimensions and conveniently mapping the coefficients to elements of the energy momentum tensor.  The angular portion is made discrete using a similar process.  A more complete description of the derivation of the expansion coefficients and the subsequent discretization of the momentum space is described in the work \cite{2011PhRvC..84c4903R}.  

\begin{figure}[h]
	\includegraphics[width=0.5\linewidth]{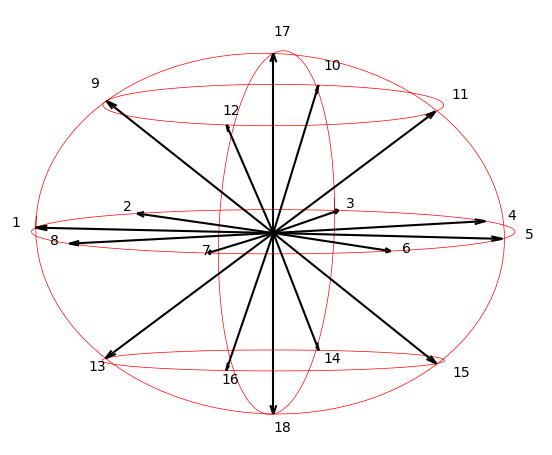}
	\caption{A sketch of a spherical lattice node in an RLBM lattice.  The momentum lattice connecting the nodes is spherical instead of box-like.}
	\label{fig:d3q19sphere}
\end{figure}

Having determined the set of orthogonal polynomials to be used to expand the probability distribution function, a discrete set of momenta can now be determined from their roots using the same quadrature technique described in (\ref{section:stdLBM}).  As the polynomials are orthogonal with respect to the solid angle, the resulting discrete three dimensional momentum lattice is spherical, and is not space filling in general (off-lattice).  This is overcome computationally using a linear interpolation scheme.  An example of a spherical lattice node is shown in figure \ref{fig:d3q19sphere}.

It remains to obtain the macroscopic moments of the fluid system from the probability distribution function by evaluating the energy-momentum tensor.  Using equation (\ref{eq:rel_pdf_exp_coef}) and (\ref{eq:energyMomentumIntegral}) we find:
\begin{equation}
T^{\mu \nu} = \frac{3 T_0^4}{ \pi^2 } 
\left(
\begin{matrix}
	a^{(0)} & a_i^{(1)} \\
	a_j^{(1)} & a_{ij}^{(2)} + \frac{1}{3} \delta^{ij} a^{(0)}
\end{matrix}
\right) 
\end{equation}
The fluid velocity $u^\mu$ is determined by the eigenvalue problem set up by the Laudau-Lifshitz condition, $u^\mu T_\mu^\nu = \epsilon u^\nu $, with the energy density as the eigenvalue $\epsilon$.  The temperature is determined from the relation $\epsilon = \frac{3 T^4}{ \pi^2 }$, and pressure $P$ is found from the ideal equation of state, $P = \frac{\epsilon}{3}$.

	\subsection{Turbulence Inducing Driving Term}
The relativistic modifications to the standard LBM having been determined, the RLBM numerical algorithm follows the standard process of (a) initialization, (b) streaming the new probability distribution function to the neighboring lattice nodes, (c) calculation of the macroscopic moments, (d) calculation of the equilibrium probability distribution function, (e) calculation of the collision term and subsequently the new probability distribution function, and (f) repeat at (b).  However, to model a turbulent flow an additional modification is required.  

A turbulent flow can be created and sustained by the introduction of a suitably large externally imposed stirring force, zero averaged in the spatial component, $ F^\mu = ( 0, F^i) $.  The force is incorporated in the RLBM algorithm in the forcing term of the force-included Relativistic Boltzmann Equation.
\begin{equation} \label{eq:boltzmanneq_force_rel}
\left[ p^\mu \nabla_\mu - \Gamma_{\mu\nu}^\lambda p^\mu p^\nu  \partial_{\lambda}^{(p)} -  F^\mu \nabla_\mu \right] f = 
- \frac{p^\mu u_\mu }{ \tau_R } \left( f - f_J^{eq}  \right)
\end{equation}
As with the probability distribution function and the equilibrium probability distribution function, the driving term must be expressed in terms of the orthogonal polynomials 
\begin{equation}
F^\mu\partial_\mu f\approx e^{-\bar{p}}\sum_{n} {a^{(n)}P^{\left(n\right)}}
\end{equation}
The orthogonality relations again determine the expansion coefficients, and are given in (\ref{eq:rel_pdf_exp_coef}). It is straightforward to show the projected form of the driving term evaluated in terms of the probability distribution function and the fluid velocity as (see appendix \ref{appendixB}):
\begin{align} \label{eq:forceTerm}
\nonumber F_i \partial_p^i f \approx
    \frac{e^{-\bar{p}}}{T_0}  & \left[  
            \frac{3}{12} F_i         \int d\chi\ v^i f
-           \frac{1}{2}  F^i v^i     \int d\chi\ f \right. \\
& +  \left. \frac{5}{4}  F_i v_j v_k \int d\chi\ f v^i v^j v^k 
-           \frac{5}{2}  F_i v_i v_k \int d\chi\ f v^k 
\right]
\end{align}
where the integration, $ \int d\chi = \int_{0}^{\infty} d\bar{p} \int \frac{d\Omega}{4\pi} $, can be evaluated using the quadrature technique described.  The integrals in (\ref{eq:forceTerm}) are then expressed as expansions with the expansion coefficients in (\ref{eq:rel_pdf_exp_coef}) up to the third order.  The driving term is calculated and applied in the same step of the LBM process as the collision calculation, and is added to the new probability distribution function, $f^* = f + \Delta t C + F^\mu \nabla_\mu f $.  With this modification in place the RLBM is equipped to produce turbulence and to model a turbulent relativistic hydrodynamic system.

    \section{Characteristics of 2D Turbulence}

When the velocity of a fluid flow reaches a critical speed relative to the viscosity of the fluid, i.e. it attains a critical Reynolds number ($Re=\frac{uL}{\nu}$ where $u$ is the fluid velocity, $L$ is the characteristic linear dimension and $\nu$ is the kinematic viscosity), in the regime of low Mach number the flow will tend to be turbulent.  An interesting and widely studied feature of turbulent flow is its tendency to create vortices at different scales within the fluid, expressed by $\mathbit{\omega} = \nabla \times \mathbit{v}$.  Due to the non-linearity of the equations of motion obeyed by the flow, the energy forming the vortices at an initial Fourier scale tends to migrate to larger and smaller scales, forming more vortices at the different scales.  This is a well established phenomenon known as the cascade of the power spectrum. The power spectrum $E(k)$ is defined as the energy density contained at a given wave number magnitude (denoted simply as $k$), or the contribution to the energy density by a wave number magnitude \cite{frisch_1995}.
\begin{equation} \nonumber
E\left( k \right) \equiv \frac{ d \mathcal{E} \left( k \right) }{dk}
\end{equation}  
$ \mathcal{E} \left( k \right) $ is the energy density in spectral space defined as a function of $k$: $ \mathcal{E} \left( k \right) \equiv \frac{1}{2} \langle | \tilde{v} (\mathbit{k}) |^2 \rangle $, where the brackets represent an average over the angular direction.  The contribution of the total energy density at all scales determines the energy of the system.
\begin{equation}  \nonumber
\mathcal{E} \left( k \right) = \int_0^{\infty} E\left( k' \right) dk'
\end{equation}

When energy is injected into the system at smaller scales (larger wave numbers), Kolmogorov \cite{1941DoSSR..30..301K} reasoned that the energy would cascade to larger scales (smaller wave numbers) at a rate proportional to the wave number at the $-5/3$rd power.  This is the so-called $-5/3$rd power law of the inverse cascade of the power spectrum \cite{1941DoSSR..30..301K}.  Specifically, Kolmogorov calculated through dimensional reasoning that the power spectrum would decay as $ \epsilon^\frac{2}{3}\ k^{-\frac{5}{3}}$ where $\epsilon$ is the energy dissipation rate.  The form of this power law was derived explicitly by Kraichnan \cite{1967PhFl...10.1417K}, and has since been observed in simulated and experimental data \cite{2001PhRvE..64c6302D}.  When energy is injected into the system externally at and/or above a driving scale $k_f$, an inverse cascade of the power spectrum is observed in the inertial range between $k_f$ and the largest scales available to the system, $ 0 < k < k_f$, and the slope of the power spectrum in that range will be $-5/3$.

It is noteworthy that a 2006 work by R. K. Scott adjusts the power law to be proportional to the wave number at power $-2$ (\cite{PhysRevE.75.046301}) in a flow with a non-zero rate of enstrophy transfer.  Then the power spectrum decays as $\epsilon^{\frac{1}{2}} \eta^{\frac{1}{6}} k^{-2} $, where here $ \eta $ is the enstrophy dissipation rate.  This spectral slope was observed in numerical studies by Westernacher-Schneider \cite{Westernacher-Schneider2017} \cite{PhysRevD.96.104054}.  However, because the forcing scheme described in the present work prevents an enstrophy transfer, Kolmogrov's slope of $-5/3$ is used as the theoretical comparison.

    \section{Model, Method and Results}
   
We consider a two-dimensional isotropic hydrodynamic flow of massless particles under a random, zero-averaged stirring force with infinite boundaries.  The RLBM numerical scheme models the system using a spherical D3Q27 momenta configuration connecting $N^2$ lattice nodes, each containing 27 discrete momenta.  The two-dimensional system is represented with a D3Q27 configuration by limiting the number of nodes in one spatial dimension to 1.  The equation of state used in the numerical scheme, that of a three dimensional ideal gas, is applicable for this two-dimensional model given the lattice configuration and the two-dimensional forcing scheme.  The boundaries of the lattice are periodic and the initial state sets the probability distribution of the momenta at each point as flat; i.e. equivalent at each node on the lattice. Simulations were conducted with lattice size $N$ set to $128, 256, 512,$ and $1024$ and the lattice spacing $\alpha$ set to $0.32, 0.16, 0.08$ and $0.04$ lattice units respectively, so that lattice configurations model systems with constant volume.  The viscosity, represented as the ratio of viscosity to entropy $\frac{\eta}{s}$, is set to $0.01$ in the collision term.    

Simulations begin initially at rest, and turbulence is introduced into the system as a response to a stirring force applied in Fourier space in zero-averaged random (spectral) directions and at a defined range of wave numbers.  A filter in Fourier space restricts the driving wave number range from a chosen wave number magnitude $k_f$ to the largest wave number magnitude permitted by the discrete Fourier lattice.  Therefore, the driving range is defined as $ k_f \leq k < k_{max} $, leaving the inertial range as $0 < k < k_f $.  Most tests were performed with $k_f$ set to $6$, and the maximum wave number $k_{max}$ varying with the lattice spacing.  At each node in the spectral lattice within the driving range, a driving term is applied at a random magnitude up to a maximum $|F|$, and in a random spectral direction within the plane so that it averages to zero throughout the lattice.  The maximum forcing magnitude is chosen to produce an adequate fluid velocity in the inertial range to induce a turbulent flow.  The forcing spectral lattice is then transformed to configuration space and applied to the relativistic lattice Boltzmann equation as a part of the forcing term.  Each simulation is run in $0.05$ time unit increments for $50.0, 100.0, 200.0$ or $400.0$ lattice time units.  Macroscopic variable calculation is conducted at each time step, and recorded every few time steps.

\begin{figure}[h]
	\begin{minipage}{0.75\textwidth} 
		\centering
		\includegraphics[width=1.0\linewidth]{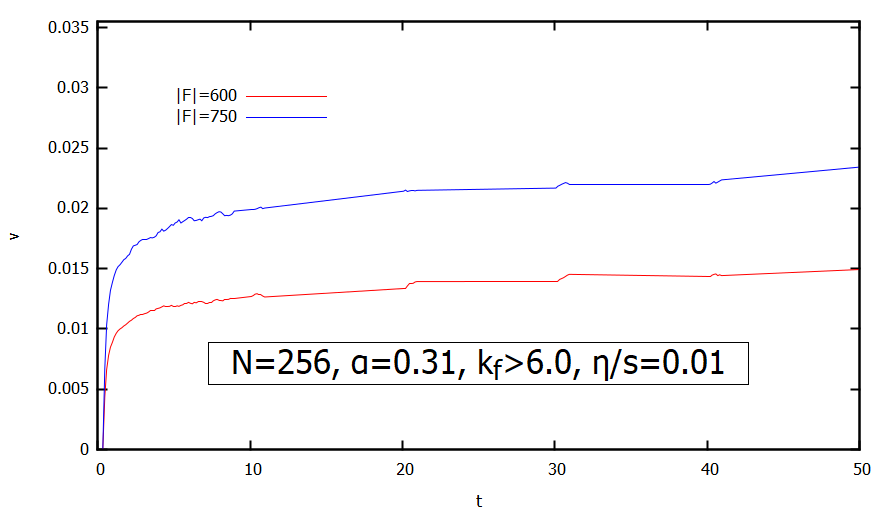}
	\end{minipage} \hfill
    \caption{Filtered average fluid velocity throughout the execution of a simulation at different maximum driving force magnitudes.  The simulations were executed with $256^2$ lattice nodes and a lattice spacing of $0.31$.  The average velocities are stable after about 32 simulated time units.}
    \label{fig:velocity-0256}
\end{figure}

\begin{figure}[h]
	\centering
	\includegraphics[width=0.75\linewidth]{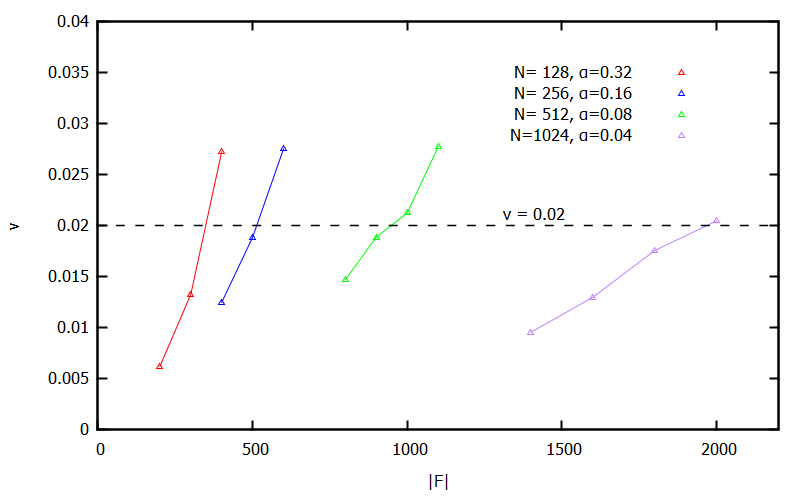}
	\caption{A constant filtered average velocity is determined across multiple lattice configurations empirically using a plot as shown.  A variation of the lattice spacing in the lattice configuration changes the number of discrete wave numbers, changing the size of the driving filter and affecting the average velocity of the simulation.  A constant velocity across multiple lattice configurations is determined by finding a common constant velocity simulated with varying maximum force magnitudes.}
	\label{fig:find-avg-velocity}
\end{figure}

The fluid velocities recorded are filtered to eliminate the contribution of the driven spectral range.  This provides a better analysis of the effects of the propagation of energy into the inertial spectral range, but note that recorded fluid velocities represent only a portion of full average fluid velocities.  The velocities obtained for each test varied based on the size of the stirring force filter and the maximum magnitude applied.  They were observed to be stable and consistent after a sufficient amount of time (Fig \ref{fig:velocity-0256}).  The stability allows the filtered average velocity to be effectively tuned as a parameter to maintain a consistent value between simulations with different configurations by selecting an appropriate stirring force maximum magnitude $|F|$.  This is done empirically by comparing the average fluid velocity of simulations with different forcing configurations for a given lattice and matching it with those of other lattices.  Figure \ref{fig:find-avg-velocity} shows an example of a line of constant velocity between four different lattice configurations with different forcing filters.    

 \hfill

For modest driving force magnitudes, the numerical model produced turbulence inducing fluid velocities of $\sim 0.01$ in the inertial range, but yielded no turbulent signal when the driving magnitude was too small.  In an inviscid system even small velocities can induce turbulence, but the model simulations suggest the existence of a minimum velocity to show energy propagation across the spectrum in a system with non-zero viscosity.  In the RLBM scheme, as viscosity approaches zero, the collision operation becomes infinitely strong (see eq. (\ref{eq:bgk_cl}) noting $\tau_R \propto \eta $), so a non-zero viscosity is required.  For a lattice spacing of $0.01$, a minimum viscosity of $0.005$ was needed to keep the evaluation of the probability distribution function numerically stable, but a smaller viscosity will not affect the stability of models with smaller lattice spacings.  The maximum viscosity was found to be $0.18$ at the same lattice spacing, also determined by stability.  Some other instabilities were observed as a result of excessively high fluid velocities present in the driven range, which are required to induce energy propagation to the inertial range.  

\begin{figure}[h]
		\centering
		\includegraphics[width=1.0\linewidth]{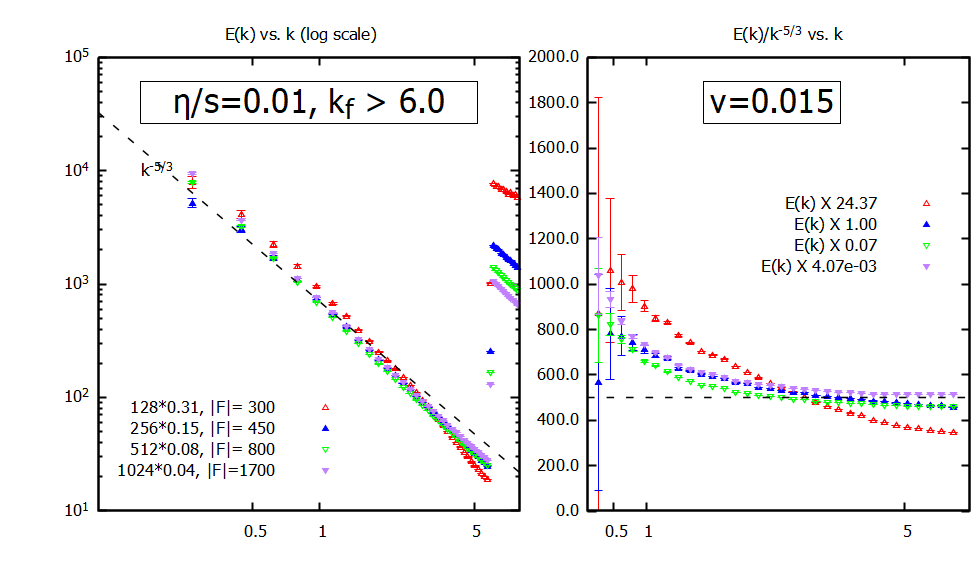}
		\caption{The energy spectrum for simulations with lattice configurations of $N\alpha = (128)(0.32), (256)(0.16), (512)(0.08)$, and  $(1024)(0.04)$, keeping a constant volume.  The energy for each lattice configuration is scaled to a similar magnitude so the curves can be easily compared.  The left panel shows the slope of the energy spectrum at log scale as compared to the expected power law of $-5/3$ in the inertial range.  Part of the driven range is visible at $k>6.0$, marked by a discontinuity in the slope to a much higher energy value.  The right panel is a "close up" of the same divided by the expected slope.  Close conformity to the $-5/3$ slope is noted, with deviations at larger scales, and for configurations with larger lattice spacing. }
		\label{fig:hiband-filter-const-vol}
\end{figure}

With adequate fluid velocity the RLBM numerical scheme shows evidence of the spectral energy propagation caused by turbulence, reflecting the slope of the energy in k-space described by Kolmogorov.  Figure [\ref{fig:hiband-filter-const-vol}] shows the power spectrum for systems modeled with different lattice configurations maintaining a constant volume.  Lattice configurations with $N$ set to $128, 256, 512$, and $1024$ are shown with lattice spacings of $0.32, 0.16, 0.08$, and $0.04$ respectively.  The driving filter is applied at $k_f = 6.0$, and at a maximum forcing magnitude determined for each to produce a constant velocity of $0.015$.   The left panel is the energy spectrum at log scale comparing the slope of the power spectrum against Kolmogorov’s predicted slope of $-5/3$.  The right panel is a “zoomed in” view of the same spectrum divided by the Kolmogorov slope in order to highlight the conformity.  The energies of the various configurations are scaled independently so that they align on the plot for comparison.
	
 \hfill

An inverse cascade was observed with a reasonable reproduction of Kolmogorov's power law of the power spectrum in the inertial range for mid and large wave numbers.  Deviations at large scales ($ \sim N\alpha $) is noted, and is presumed to be a product of the limited discrete spectrum causing a “pileup” at the largest discrete scales.  The effect is eased by changing the lattice spacing to create more small $k$ modes below the driving range, giving the energy more modes to which it can propagate.  Because of the larger errors, the results at that scale of the system are not considered in the conclusions.  The simulation configured with $N = 128$ and $\alpha=0.32$ shows a significant pileup at the large scales and lower conformity to the $-5/3$ power law at small scales.  The smaller number of nodes requires a larger lattice spacing to maintain a the same  volume, which maximizes the distortion of a discrete lattice and impedes its ability to accurately model a continuous system.  

\begin{figure}[h]
		\centering
		\includegraphics[width=1.0\linewidth]{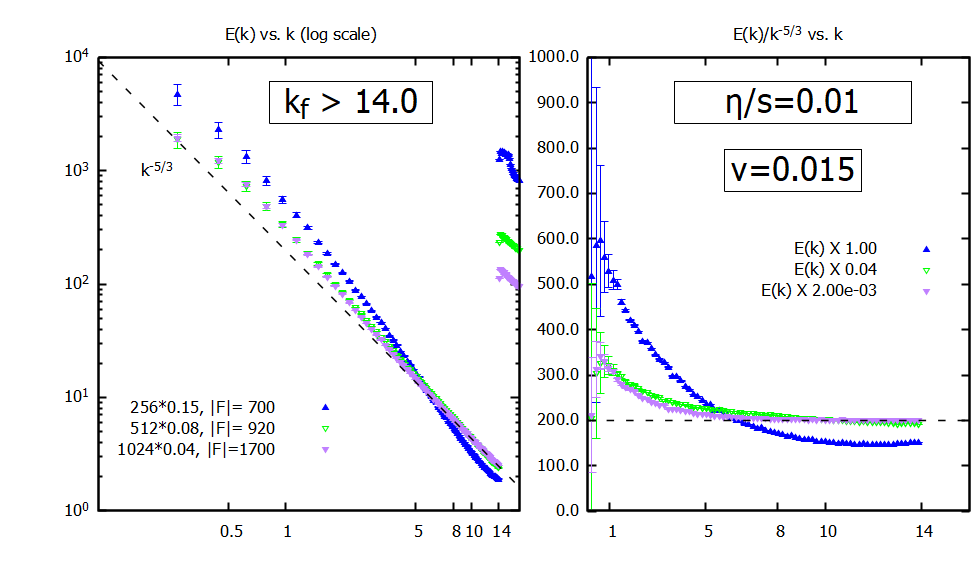}
	\caption{The energy spectrum of the inertial range $0 < k < 14.0$ for simulations with lattice configurations of $N\alpha = (256)(0.16), (512)(0.08)$, and $(1024)(0.04)$, scaled for comparison.  The left panel shows the slope of the energy spectrum at log scale as compared to the expected power law of $-5/3$, and the right panel is a "close up" of the same divided by the expected slope.  A larger inertial range shows improved conformity.}
	\label{fig:hiband-kf14-constVol}
\end{figure}

As the inertial range of the spectrum expands the power spectrum forms a better resemblance to the $-5/3$ power law and the distortion at the large scales is eased, implying a closer conformity at the continuum limit.  That is, as the number and range of discrete wave numbers in the inertial range grows the slope of the power spectrum approaches a closer resemblance to what is expected in a continuous system.  Figure \ref{fig:hiband-kf14-constVol} compares the same lattice configurations as in figure \ref{fig:hiband-filter-const-vol}, but driven with a forcing configuration leaving the inertial range from $k = 0$ to $k = 14$.  This leaves the $256^2$ node configuration simulation driving at only a third of the accessible wave numbers which disrupted the spectral slope.  But the spectra of the larger lattice configurations show a much better conformity to the power law in a larger portion of the inertial range.  Simulations also showed that the larger the inertial range, the more kinetic energy injected into the system is required to achieve the energy propagation.  Indeed, a fluid velocity of $\approx 0.25$ was stable for a $N\alpha = (1024)(0.04)$ configuration with a forcing configuration of $k_f=60$ and $|F|=6000$, where the driven range is only 30\% of the available spectrum.  

\begin{figure}[h]   
		\centering
		\includegraphics[width=0.75\linewidth]{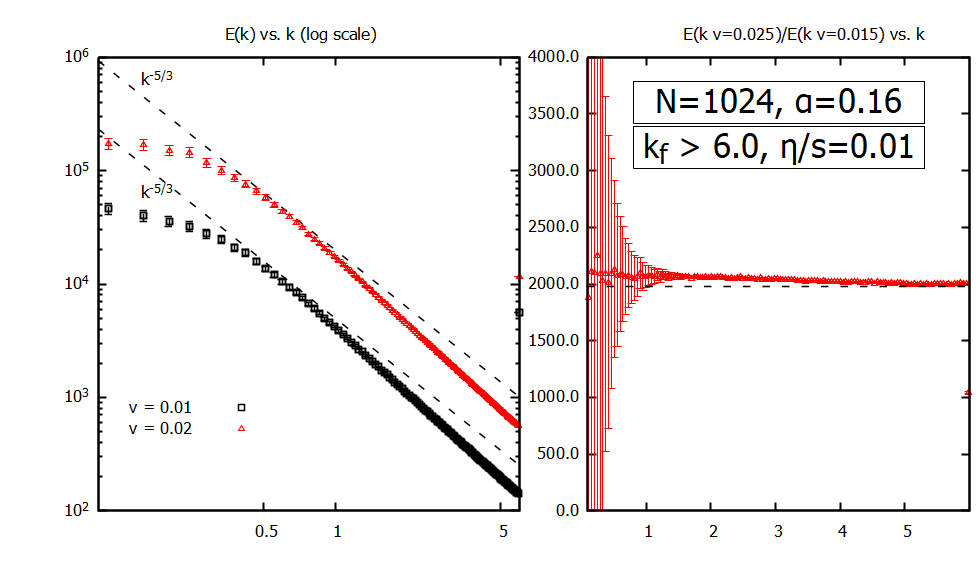}
	\caption{The energy spectrum of a simulation with lattice configuration $N\alpha = (1024)(0.16)$ with large and small filtered average velocity.  The left panel compares the energy spectrum at log scale with the expected slope of $-5/3$.  The right panel is the energy spectrum of the larger velocity simulation divided by the smaller velocity simulation.  The magnitude of the fluid velocity does not have an effect on the inverse energy cascade, provided the velocity is sufficient to produce turbulence.}
	\label{fig:hiband-hilo-multiN}
\end{figure}

The energy spectrum of a larger velocity flow is compared to a smaller velocity flow for a $ N\alpha = (1024)(0.16)$ lattice configuration in figure [\ref{fig:hiband-hilo-multiN}].  The left panel is the energy spectrum at log scale, and the right panel is the energy spectrum of the larger velocity system divided by the smaller one which is rendered as a zero-slope line.  No discernible differences in the energy spectrum’s slope due to velocity is detected, providing evidence that the numerical model's energy propagation is not dependent on the average velocity, provided it is significant enough induce energy propagation.

\begin{figure}[h] 
	\centering
	\includegraphics[width=0.75\linewidth]{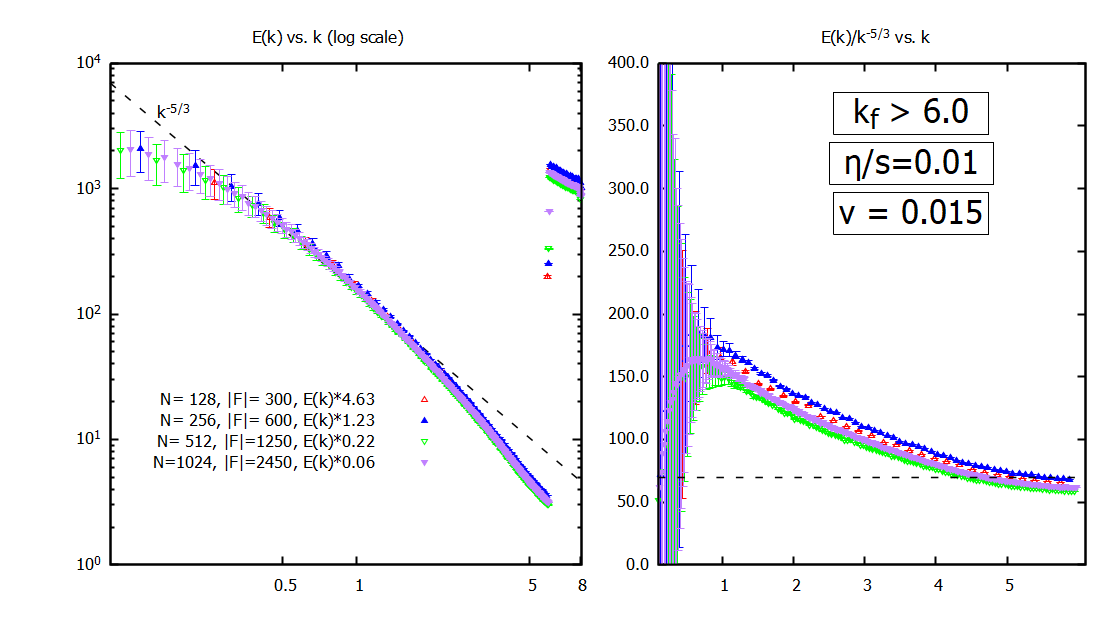}
	\caption{The energy spectrum of lattice configurations $N\alpha = (128)(0.16), (256)(0.16), (512)(0.16)$, and $(1024)(0.16)$ (increasing volume), with a constant filtered average fluid velocity of $0.015$.  The left panel compares the energy spectrum to the expected slope of $-5/3$ at log scale, and the right is the energy spectrum divided by the expected slope.  The energy spectrum in each panel is scaled for comfortable comparison.  The similarity of the slope of each curve suggests the energy propagation is not sensitive to volume when the volume is controlled by the number of lattice nodes.}
	\label{fig:hiband-filter-incr-vol}
\end{figure}

\begin{figure}[h]
	\begin{minipage}{0.48\textwidth}
		\centering
		\includegraphics[width=1.0\linewidth]{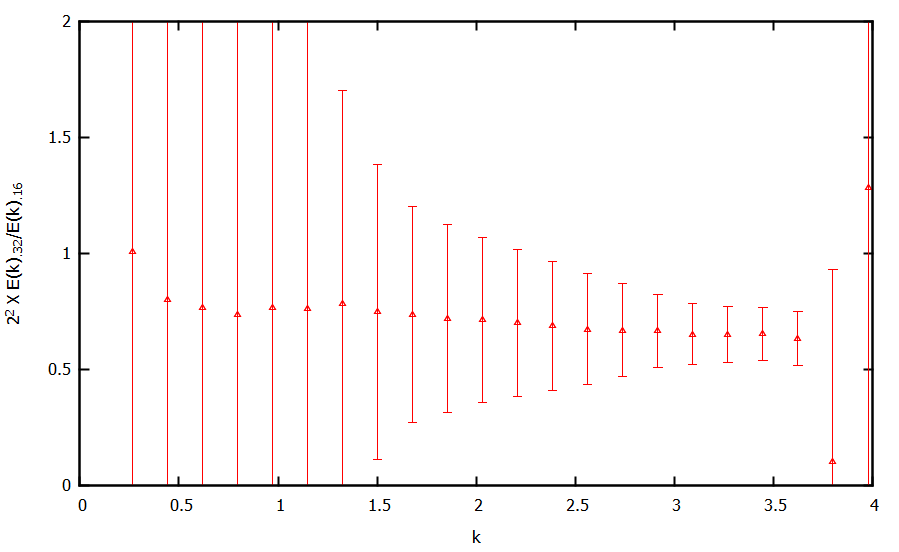}
		\label{fig:ratio-latspc-diffvol}
	\end{minipage}
	\begin{minipage}{0.48\textwidth}
		\centering
		\includegraphics[width=1.0\linewidth]{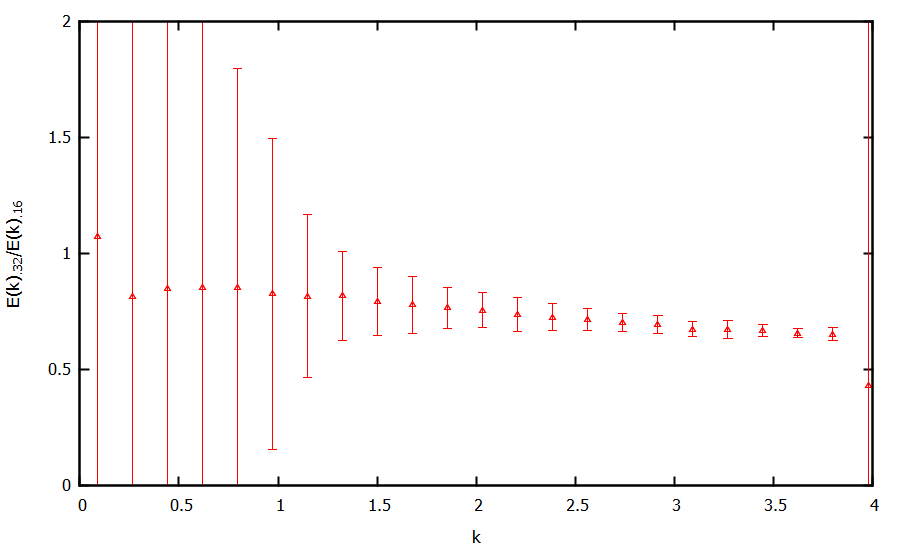}
		\label{fig:ratio-latspc-samevol}
	\end{minipage}
		\caption{ Left is the ratio of the power spectrum $E(k)$ for simulations with lattice spacing $0.32$ and $0.16$ and a constant $N$.  The ratio at each $k$ trends to constant, suggesting the simulated results are not affected by lattice spacing alone. The $\sim \frac{3}{4}$ ratio is attributed to an additional mode available with a tighter lattice. The lattice with the smaller volume is scaled with a factor of $2^2$ to compensate for the larger energy capacity of the larger volume system.  Right is the same power spectrum ratio for simulations with lattice configurations $N\alpha = (128)(0.032)$ and $(128)(0.016)$.  The ratio also tends to be constant, suggesting the simulated results are not affected by the non-physical lattice parameters $N$ and $\alpha$.}
		\label{fig:ratio-latspc-diffsamevol}
\end{figure}

Figure [\ref{fig:hiband-filter-incr-vol}] shows the energy spectrum for systems of increasing size, modeled by an increasing number of lattice nodes with constant lattice spacing.  The conformity to the expected slope mimics the simulations performed with a constant volume implying the model's propagation of energy is not dependent on the number of nodes, a non-physical parameter.  Figure (\ref{fig:ratio-latspc-diffsamevol}) (left) shows the scaled ratio of $E(k)$ observed for a simulation with a lattice spacing of $0.32$ to $E(k)$ for a simulation with a lattice spacing of $0.16$.  The absence of a slope for the ratio suggests the power spectrum is not sensitive to lattice spacing alone.  A similar comparison (right) of the energy spectrum for systems with a constant volume yields a similar ratio.  Finally, the energy propagation was found to be independent of the configuration of the stirring force (the choice of $k_f$, effective $k_{max}$, and $|F|$) provided it is large enough to induce a sufficient velocity capable of producing turbulence.

    \section{Conclusions}

To explore the properties of a massless fluid system we have incorporated a stirring force into a relativistic modification of the Lattice Boltzmann Method, driving the simulation of a two-dimensional turbulent fluid system at small spectral scales.  The energy spectrum of the modeled flow demonstrates agreement with Kolmogorov’s predicted slope of $-5/3$ in the inertial range indicating an inverse cascade of energy expected for a turbulent flow.  The error is large at large scales and shows evidence of a pileup, particularly for lattice configurations with a smaller number of scales in the infrared range.  But the slope is reasonably reproduced in the inertial range above the large scales up to the driven scales, implying the model is a reasonable reproduction of turbulent flows for a massless fluid.  The model is shown to be insensitive to non-physical parameters including lattice size except with respect to the effects of a large lattice spacing.  The model is readily applied to a two dimensional system such as graphene or a Kagome solid, where the charge carriers flow as a relativistic fluid and relativistic hydrodynamics is applicable.  Within a Kagome solid in particular, a turbulent massless fluid flow is thought to be within reach of experiment.  Then the flow characteristics investigated with this model can be demonstrated empirically.  Further applications of the RLBM in the exploration of turbulent flows should continue to be investigated.  In particular, a three dimensional model should be tested, and different turbulence generating schemes should be considered in order to achieve higher velocities.  

We thank P. Romatschke for his instruction, helpful discussions and for access to an RLBM source code.  We'd also like to thank John Ryan Westernacher-Schneider and Luciano Rezzolla for their helpful comments and insight.  This work was supported by the Department of Energy, DOE award No DE-SC0017905.  All simulations were run on the Eridanus cluster at the University of Colorado Boulder.

\appendix
	
\section{Three Dimensional Orthogonal Polynomials} \label{appendixA}	

The vector polynomials $ P_{i_1 ... i_n}^{(n)} \left( \mathbit{v} \right) $ are constructed through the orthogonality condition with respect to the angular integral $ \int \frac{d\Omega}{4\pi} $.  The first few polynomials are 
\begin{align}
\nonumber P^{(0)} =& 1, \\
\nonumber P_i^{(1)} =& v_i, \\
\nonumber P_{ij}^{(2)} =& v_i v_j - \frac{1}{3} \delta_{ij} , \\
\nonumber P_{ijk}^{(3)} =& v_i v_j v_k - \frac{1}{5} \left(  \delta_{ij} v_k + \delta_{ik} v_j + \delta_{jk} v_i \right), \\
...
\end{align}
The first few orthogonality relations are
\begin{align}
\nonumber \int \frac{d\Omega}{4\pi} P^{(0)} P^{(0)} = & 1, \\
\nonumber \int \frac{d\Omega}{4\pi} P_i^{(1)} P_j^{(1)} = & \frac{\delta_{ij}}{3} , \\
\nonumber \int \frac{d\Omega}{4\pi} P_{ij}^{(2)} P_{lm}^{(2)} = & \frac{1}{15} \left( \delta_{il}\delta_{jm} + \delta_{im}\delta_{jl} - \frac{2}{3} \delta_{ij}\delta_{lm}  \right)
\end{align}

\section{Discrete the Forcing Term} \label{appendixB}

The forcing term is projected onto orthogonal polynomials $P^{(n)}$ up to the second order. 
$$
F_i (\partial_p^i f) = e^{-\bar{p}} \sum_{n=0}^{2} {a^{(n)}P^{(n)} }
$$
The projection coefficients $a^{(n)}$ are dependent on position and time, and the external force $F_i$ is independent of momentum.  The divergence of the Boltzmann probability distribution function $f$ is with respect to momentum.  The time element of momentum, $\bar{p}=\frac{p^0}{T_0}$, is also the magnitude of the momentum.  The first three multi-dimensional orthogonal polynomials are:
$$
P^{(0)}(\mathbit{v})=1, \ \  P_i^{(1)}(\mathbit{v})=v_i, \ \  P_{ij}^{(2)}(\mathbit{v})=v_iv_j - \frac{1}{3}\delta_{ij} .
$$

The coefficients are obtained with the orthogonality relationship,  
\begin{eqnarray}
\nonumber
\int_{0}^{\infty} d\bar{p} \bar{p}^2 \int{\frac{d\Omega}{4\pi} {\bar{p}} F_i \left( \partial_p^if \right) P^{(m)}} &=& 
\int_{0}^{\infty} d\bar{p} \bar{p}^2 \int{\frac{d\Omega}{4\pi} {\bar{p}} \left( e^{-\bar{p}}\sum_{n}^{\infty}{a^{(n)}P^{(n)}} \right)P^{(m)}}  \\
\nonumber
&=& 
\int_{0}^{\infty} d\bar{p} \bar{p}^2 \int{\frac{d\Omega}{4\pi} {\bar{p}} e^{-\bar{p}}a^{(m)}\left(P^{(m)}\right)^2} .
\end{eqnarray} 

For the zero order coefficient, the orthogonality relationship is,
\begin{eqnarray}
\nonumber
\int_{0}^{\infty} \int{\frac{d\Omega}{4\pi} } d\bar{p}\ {\bar{p}}^3\ F_i \left(\partial_p^i f\right) P^{(0)} &=& \int_{0}^{\infty} \int\frac{d\Omega}{4\pi} d\bar{p}\ {\bar{p}}^3 e^{-\bar{p}} a^{(0)} \left(P^{(0)}\right)^2 .
\end{eqnarray}
The integrals are regarded as spherical volume integrals in velocity space where $\Omega$ is the solid angle, and $\bar{p}$ is the unit-less magnitude of the momentum representing the radius extending to infinity.  We obtain,
\begin{equation}
\label{eq:expCoef01}
a^{(0)} = - \frac{F_i}{6 T_0} \int_{V} dV v^i f .
\end{equation}
where the volume integral is represented as $\int_{V} dV = \int_{0}^{\infty} d\bar{p}\ {\bar{p}}^2 \int \frac{d\Omega}{4\pi}$.  The divergence theorem at an infinite boundary was used, and the divergence of the magnitude of the momentum, given by $\partial_p^i \bar{p} = \frac{v^i}{T_0}$, was also used.  

The orthogonal expression for the first order coefficient is, 
\begin{eqnarray}
\nonumber
\int_{0}^{\infty} \int \frac{d\Omega}{4\pi} d\bar{p}\ {\bar{p}}^3\ F_i \left(\partial_p^if\right) P^{(1)j} &=& \int_{0}^{\infty} \int \frac{d\Omega}{4\pi} d\bar{p}\ {\bar{p}}^3 e^{-\bar{p}} a_k^{(1)} P^{(1)k} P^{(1)j} , \\
\nonumber
\int_{0}^{\infty} \int \frac{d\Omega}{4\pi} d\bar{p} {\bar{p}}^3 F_i \left(\partial_p^if\right) v^j &=& 2 a_j^{(1)} , \\
\nonumber
- F_j \frac{1}{T_0} \int_{V} dV f &=& 2 a_j^{(1)} , \\
\label{eq:forceExpCoef00}
a_j^{(1)} &=&  - \frac{1}{2} \frac{F^j}{T_0} \int_{V} dV f .
\end{eqnarray}
where we evaluated the angular portion of the integral with the relationship $\int \frac{d\Omega}{4\pi} v^i v^j = \frac{1}{3} \delta_{ij}$, and we used the divergence of the velocity relation $\partial_p^i v^j = \frac{1}{p^0} ( \delta^{ij} - v^i v^j ) $.

Finally, we used the angular integral $\int \frac{d\Omega}{4\pi} v^i v^j v^l v^m = \frac{1}{15} \left( \delta_{il}\delta_{jm} + \delta_{im}\delta_{jl} + \delta_{ij}\delta_{lm} \right) $ to determine the second order projection coefficient.
\begin{eqnarray}
\nonumber
\int_{0}^{\infty} \int \frac{d\Omega}{4\pi} d\bar{p}\ {\bar{p}}^3\ F_k \left( \partial_p^k f\right ) P^{(2)lm} &=& \int_{0}^{\infty} \int \frac{d\Omega}{4\pi} d\bar{p}\ {\bar{p}}^3 e^{-\bar{p}} a_{ij}^{(2)} P^{(2)ij} P^{(2)lm} , \\
\nonumber
a_{ij}^{(2)} \frac{4}{5} & = &
F_k \frac{1}{T_0} \int_{V}dV f v^k v^i v^j 
- 2 F_i \frac{1}{T_0} \int_{V}dV f v^j \\ \nonumber
& \ & +\frac{1}{3} \delta^{ij} F_k \frac{1}{T_0} \int_{V} dV f v^k    
\end{eqnarray}

For this calculation we note the trace of the velocity is unity, containing only angular information.  Also, the second order orthogonal polynomial is trace-less, as is its coefficient  $a_{ij}^{(2)}$, so that $ a_{ij}^{(2)} \delta_{ij} = 0 $.  Then the second order projection coefficient is,

\begin{equation}
\label{eq:forceExpCoef02}
	a_{ij}^{(2)}  
	=   \frac{5}{4}  F_k \frac{1}{T_0} \int_{V}dV f v^k v^i v^j
	- \frac{5}{2}  F_i \frac{1}{T_0} \int_{V}dV f v^j 
	+ \frac{5}{12} F_k \frac{1}{T_0} \delta^{ij} \int_{V} dV f v^k .
\end{equation}

The forcing term projected onto orthogonal polynomials up to the second order is then, 
\begin{align}
	\nonumber
	F_i \partial_p^i f 
	\approx e^{-\bar{p}}  \frac{1}{T_0}  & \left[
	          \frac{3}{12} F_i \int_{V}dV v^i f  
	-         \frac{1}{2}  F^i v^i \int_{V} dV f \right. \\ 
&	+ \left.  \frac{5}{4}  F_i v_j v_k \int_{V}dV f v^i v^j v^k
	-         \frac{5}{2}  F_i v_i v_k \int_{V}dV f v^k 
	\right] .
\end{align}

\section{Minimum Simulation Time} 

\begin{figure}[h] 
	\centering
	\includegraphics[width=0.75\linewidth]{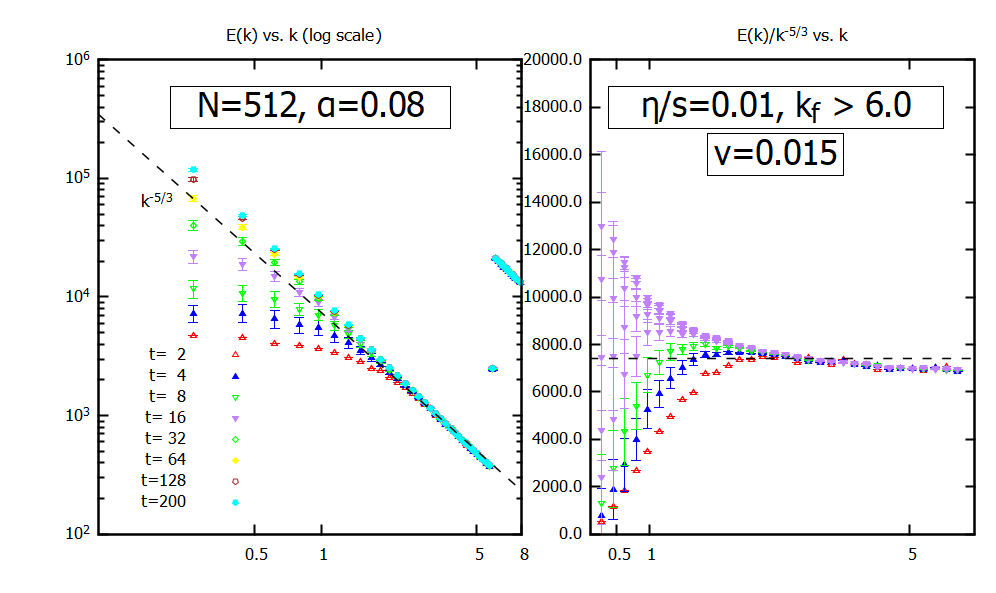}
	\caption{Power spectrum for simulations with energy injected at $k_f = 6.0$ at driving force magnitude $|F|=800$ for simulation durations $t=2$ through $200$.  A general conformity to the Kolmogorov spectrum is obtained after $t=32$.}
	\label{fig:mintimedualplot}
\end{figure}

Convergence to a stable energy spectrum for a simulation was achieved before $32$ simulated time units.  Figure (\ref{fig:mintimedualplot}) shows the power spectrum for a simulation at $t=2$ to $200$ time units and indicates a stable spectral slope was realized after $t=32$ and remained stable to large time windows.  The companion plot depicts the same values divided by the predicted slope providing a higher resolution picture.  The plot depicts large errors at the large scales; a result of numerical limitations for small wave numbers in a discrete spectrum.  However, the error for the spectrum for the shorter time span engulfs the spectrum for longer time spans, displaying an evolving stability.

\begin{figure}[h] 
	\centering
	\includegraphics[width=1.0\linewidth]{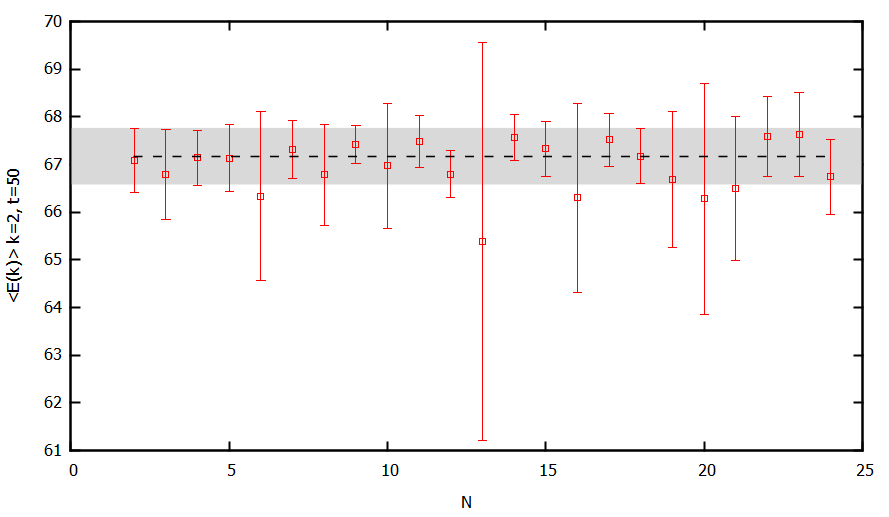}
	\caption{ The average Power Spectrum $ \langle E(k) \rangle $ vs. $N$, where $N=\frac{t}{\Delta t = 0.05}$ is a range of candidate time steps to best mitigate autocorrelation risk.  The gray band indicates agreement within the margin of error of the $N=18$ time step.  The power spectrum is for $k = 2.0$ in a simulation with energy injected at scales greater than $4.0$ at maximum driving force magnitude of $400$.  The simulation was run for $t=50$ time units. }
	\label{fig:nvsEkwithErrCT}
\end{figure}

The error is calculated as $ \frac{ \sum_{i=1}^{N} \sqrt{ \langle E(k) \rangle^2-E_N(k)^2}}{N} $ where $ \langle E(k) \rangle $ is a running average in time for each $k$.  Here $N$ is a time step determined by a suitable disassociation time $t$ where $N=\frac{t}{\Delta t = 0.05}$.  At a smaller time step the average poses an autocorrelation danger, so a suitably large $N$ was determined to remove the autocorrelation risk and maximize the number of time steps participating in the average.  Figure (\ref{fig:nvsEkwithErrCT}) shows $\langle E\left(k\right) \rangle $ for $k_f=2.0$ at $t=50$, plotted against candidate values of $N$, with the error in red.  $N = 18$ (and therefore a disassociation time of $t=0.9$) was determined to be acceptably large because of its agreement, within the margin of error, with all other values of $N$ (as shown by the gray band), and because of its relatively small error.  Many other choices of $N$, including $4$ or $5$, could have also been chosen to achieve the same result.  This analysis is not statistically rigid, but suitable enough for the purpose.

\pagebreak

\bibliographystyle{hunsrt}
\bibliography{lambda}
\end{document}